\documentclass[aps,physrev,twocolumn,superscriptaddress,nofootinbib,10pt]{revtex4-2}

\usepackage[utf8]{inputenc}
\usepackage{tabularx}
\usepackage{multirow}
\usepackage{url}
\usepackage{amsmath}
\usepackage{amssymb}
\usepackage{graphicx}
\usepackage{mathrsfs}
\usepackage{color}
\usepackage{bm}
\usepackage[normalem]{ulem}
\usepackage{float}
\usepackage{lmodern}
\usepackage{soul}
\usepackage[breaklinks,colorlinks,urlcolor=blue,citecolor=blue]{hyperref}

\usepackage{amsmath}
\usepackage{mathrsfs}
\allowdisplaybreaks

\newcommand{\araa}{ARA\&A}%
\newcommand{\apjl}{Astrophys.~J.}%
\newcommand{\mnras}{MNRAS}%

\def\simlt{\lower.5ex\hbox{$\; \buildrel < \over \sim \;$}}
\def\simgt{\lower.5ex\hbox{$\; \buildrel > \over \sim \;$}}

\def\pa{\partial}

\newbox\grsign \setbox\grsign=\hbox{$>$} \newdimen\grdimen \grdimen=\ht\grsign
\newbox\simlessbox \newbox\simgreatbox \newbox\simpropbox
\setbox\simgreatbox=\hbox{\raise.5ex\hbox{$>$}\llap
     {\lower.5ex\hbox{$\sim$}}}\ht1=\grdimen\dp1=0pt
\setbox\simlessbox=\hbox{\raise.5ex\hbox{$<$}\llap
     {\lower.5ex\hbox{$\sim$}}}\ht2=\grdimen\dp2=0pt
\setbox\simpropbox=\hbox{\raise.5ex\hbox{$\propto$}\llap
     {\lower.5ex\hbox{$\sim$}}}\ht2=\grdimen\dp2=0pt
\def\simgt{\mathrel{\copy\simgreatbox}}
\def\simlt{\mathrel{\copy\simlessbox}}

\begin{document}

\title{Propagation of strong electromagnetic waves in tenuous plasmas}

\author{Emanuele Sobacchi}
\email{emanuele.sobacchi@gssi.it}
\affiliation{Gran Sasso Science Institute, viale F.~Crispi 7, L’Aquila, 67100, Italy}
\affiliation{INAF -- Osservatorio Astronomico di Brera, via E.~Bianchi 46, Merate, 23807, Italy}
\author{Masanori Iwamoto}
\affiliation{Yukawa Institute for Theoretical Physics, Kyoto University, Kitashirakawa-Oiwakecho, Sakyo-Ku, Kyoto, 606-8502, Japan}
\affiliation{Faculty of Engineering Sciences, Kyushu University, 6-1, Kasuga-koen, Kasuga, Fukuoka, 816-8580, Japan}
\author{Lorenzo Sironi}
\affiliation{Department of Astronomy and Columbia Astrophysics Laboratory, Columbia University, 550 W 120th St, New York, NY 10027, USA}
\affiliation{Center for Computational Astrophysics, Flatiron Institute, 162 5th Avenue, New York, NY 10010, USA}
\author{Tsvi Piran}
\affiliation{Racah Institute for Physics, The Hebrew University, Jerusalem, 91904, Israel}

\begin{abstract}
We study the propagation of electromagnetic waves in tenuous plasmas, where the wave frequency, $\omega_0$, is much larger than the plasma frequency, $\omega_{\rm P}$. We show that in pair plasmas nonlinear effects are weak for $a_0 \ll \omega_0/\omega_{\rm P}$, where $a_0$ is the wave strength parameter. In electron-proton plasmas a more restrictive condition must be satisfied, namely either $a_0\ll 1/\omega_{\rm P}\tau_0$, where $\tau_0$ is the duration of the radiation pulse, or $a_0\ll 1$. We derive the equations that govern the evolution of the pulse in the weakly nonlinear regime. Our results have important implications for the modeling of fast radio bursts. We argue that:~(i) Millisecond duration bursts with a smooth profile must be produced in a proton-free environment, where nonlinear effects are weaker.~(ii) Propagation through an electron-proton plasma near the source can imprint a sub-microsecond variability on the burst profile.
\end{abstract}

\maketitle

\section{Introduction}

Understanding the propagation of electromagnetic waves in plasmas is important for both laser physics and astrophysics \cite{Kruer2019, Ginzburg1970, RybickiLightman1979}. The advent of high-power laser facilities, and the discovery of bright extragalactic radio transients, i.e.~fast radio bursts, makes it crucial to consider the regime in which electrons oscillate with relativistic velocities in the field of the wave \cite{Mourou+2006, Esarey+2009, Lyubarsky2021, Zhang2023}. In this regime, the wave strength parameter is $a_0\gg 1$ (where $a_0$ is defined as the peak transverse component of the electron four-velocity in units of the speed of light).

The propagation of electromagnetic waves in plasmas is governed by Maxwell's and two-fluid equations  \cite{Kruer2019, Ginzburg1970, RybickiLightman1979}. These equations contain nonlinear terms. In the nonrelativistic limit $a_0\ll 1$, nonlinear terms can be treated as a small perturbation by making an expansion in powers of $a_0$ \cite{Kruer2019}. This makes it possible to study the wave propagation analytically in the limit $a_0\ll 1$. However, a systematic analytic treatment of the relativistic regime $a_0\gg 1$ has been lacking so far.

In this paper, we develop a framework to study the propagation of electromagnetic waves in the regime $a_0\gg 1$. We show that nonlinear terms can be treated as a small perturbation when the plasma is sufficiently tenuous, as expected in astrophysical systems where the frequency of radio waves can exceed the plasma frequency by several orders of magnitude. We apply our framework to fast radio bursts, and discuss the imprint of the plasma surrounding the source on their time structure.

The paper is organized as follows. In Sec.~\ref{sec:fundamental} we present the fundamental equations that govern the propagation of strong electromagnetic waves. We study the fast oscillations of the physical quantities on the time scale of the wave period, and their secular evolution on much longer time scales. In Sec.~\ref{sec:envelope} we derive the equation that governs the evolution of the wave envelope. We present exact analytical solutions where the wave intensity is constant, and study their stability. In Sec.~\ref{sec:disc} we discuss the implications of our results for fast radio bursts, and summarize our conclusions.

\section{Fundamental equations}
\label{sec:fundamental}

We consider a quasi-monochromatic wave packet of arbitrary polarization that propagates in a cold unmagnetized plasma. In the frame where the particles ahead of the wave packet are at rest (hereafter, the ``lab frame''), the wave frequency is $\omega_0$, and the wave vector is $k_0\bm{e}_z$. We work in units where the speed of light is $c=1$.

The electromagnetic fields can be presented as $\bm{E}=-(m/|e|)(\nabla\phi+\pa\bm{a}/\pa t)$ and $\bm{B}=(m/|e|)\nabla\times\bm{a}$, where $m$ is the electron mass, and $e$ is the charge. We work in the Coulomb gauge, namely $\nabla\cdot\bm{a}=0$. The equation of motion can be presented as \cite{BourdierFortin1979}
\begin{align}
\nonumber
\frac{\pa}{\pa t} & \left(\bm{u}_\pm \pm \eta_\pm\bm{a} \right) + \\
\label{eq:motion}
& - \frac{\bm{ u}_\pm}{\gamma_\pm} \times \left[\nabla\times\left(\bm{u}_\pm\pm\eta_\pm\bm{a}\right)\right] = -\nabla\left(\gamma_\pm\pm\eta_\pm\phi\right)\;,
\end{align}
where $\bm{ u}_+$ and $\bm{u}_-$ are the ion and electron four-velocities, and $\gamma_\pm=\sqrt{1+\bm{ u}_\pm^2}$ is the Lorentz factor. We defined $\eta_-=1$ and $\eta_+=m/M$, where $M$ is ion mass. We consider both pair plasmas ($\eta_+=1$), and electron-proton plasmas ($\eta_+\ll 1$). The continuity equation is
\begin{equation}
\label{eq:cont}
\frac{\pa}{\pa t}\left(\gamma_\pm n_\pm\right) + \nabla\cdot\left(n_\pm\bm{ u}_\pm\right) = 0 \;,
\end{equation}
where $n_\pm$ is the proper number density. We adopt the most natural definition of the plasma frequency, namely $\omega_{\rm P}=\sqrt{4\pi(1+\eta_+)n_0e^2/m}$, where $n_0$ is the electron proper density ahead of the wave packet. The electromagnetic fields are governed by Gauss's law, which can be presented as
\begin{equation}
\label{eq:maxwell0}
\nabla^2\phi = -\frac{\omega_{\rm P}^2}{1+\eta_+} \left(\frac{\gamma_+n_+}{n_0}-\frac{\gamma_-n_-}{n_0}\right) \;,
\end{equation}
and by Amp\`ere's law, which can be presented as
\begin{equation}
\label{eq:maxwell1}
\frac{\pa^2\bm{a}}{\pa t^2} -\nabla^2\bm{a} +\nabla\frac{\pa\phi}{\pa t} = \frac{\omega_{\rm P}^2}{1+\eta_+} \left(\frac{n_+\bm{ u}_+}{n_0} - \frac{n_- \bm{ u}_-}{n_0} \right) \;.
\end{equation}

We work in the frame that moves with the group velocity $k_0/\omega_0$ (hereafter, the ``wave frame''). As discussed by Clemmow \cite{Clemmow1974}, in this frame the wave vector vanishes, and the wave frequency is $\omega=\sqrt{\omega_0^2-k_0^2}$. We solve Eqs.~\eqref{eq:motion}-\eqref{eq:maxwell1} by separating the fast oscillations of the physical quantities, which occur on the time scale $1/\omega$, from their secular evolution, which occurs on much longer time scales.

Before entering the details of the calculation, we anticipate some important facts. Since the wave vector vanishes, the spatial derivatives do not affect the fast oscillations. Then, Eqs.~\eqref{eq:motion}-\eqref{eq:cont} give $\bm{u}_\pm\simeq \mp\eta_\pm\bm{a}$, and $\gamma_\pm n_\pm\simeq$ const, showing that the current $n_\pm \bm{ u}_\pm$ is proportional to $\bm{a}/\gamma_\pm$. Particles ahead of the wave packet have a large Lorentz factor, $\gamma_0=\omega_0/\omega$. We focus on the regime in which the particle Lorentz factor inside the wave packet is nearly constant (i.e.,~$\gamma_\pm \simeq\gamma_0$). In this regime nonlinear effects are weak, as the current becomes proportional to $\bm{a}$ (when $\gamma_\pm \simeq\gamma_0$, one has $n_\pm\simeq n_0$, from which it follows that $n_\pm\bm{u}_\pm\simeq \mp\eta_\pm n_0\bm{a}$). Since the spatial derivatives do not affect the fast oscillations, the left hand side of Eq.~\eqref{eq:maxwell1} can be approximated as $-\omega^2\bm{a}$. Substituting $n_\pm\bm{u}_\pm\simeq \mp\eta_\pm n_0\bm{a}$ into the right hand side of Eq.~\eqref{eq:maxwell1}, one finds the dispersion relation of the wave, which is $\omega^2=\omega_{\rm P}^2$, or equivalently $\omega_0^2-k_0^2=\omega_{\rm P}^2$ in the lab frame. This implies $\gamma_0=\omega_0/\omega_{\rm P}$. We will show that the key condition $\gamma_\pm \simeq\gamma_0$ can be satisfied even in the regime $a_0\gg 1$ when the plasma is sufficiently tenuous.

\subsection{Fast oscillations}
\label{sec:fast}

When nonlinear effects are weak, high-order harmonics of the fields can be neglected. Then, the vector potential can be presented as
\begin{equation}
\label{eq:a}
\bm{a} = \bm{\psi}_0 + \frac{1}{2}\bm{a}_0 {\rm e}^{-{\rm i}\omega t} + {\rm c.c.}\;,
\end{equation}
where c.c.~is the complex conjugate of the fast oscillating term proportional to ${\rm e}^{-{\rm i}\omega t}$. The complex function $\bm{a}_0$ describes the envelope of the wave packet. The real function $\bm{\psi}_0$ describes the secular evolution of the vector potential.

The fast oscillations of the four-velocity can be determined by substituting Eq.~\eqref{eq:a} into Eq.~\eqref{eq:motion}.  The spatial derivatives can be neglected if $\nabla |\bm{a}_0| \ll \omega a_0$ in the wave frame. Since we will find that $\omega=\omega_{\rm P}$, the longitudinal size of the wave packet in the lab frame should be longer than a few wavelengths, and the transverse size should be longer than the plasma skin depth (these conditions are satisfied by radio bursts of GHz frequency and millisecond duration that illuminate a substantial fraction of the solid angle). Then, one finds
\begin{equation}
\label{eq:ufast}
\bm{ u}_\pm = - u_0\bm{ e}_z + \delta \bm{ u}_{0\pm} \mp \frac{1}{2} \eta_\pm \bm{a}_0 {\rm e}^{-{\rm i}\omega t} + {\rm c.c.} \;,
\end{equation}
where $u_0 = \sqrt{\gamma_0^2-1} = k_0/\omega$ is the four-velocity ahead of the wave packet, and $\delta \bm{ u}_{0\pm}$ describes the secular evolution of the four-velocity. The wave strength parameter, which is defined as the peak transverse component of $\bm{ u}_-$ in units of the speed of light, is given by $a_0=\max[|\bm{a}_0|]$.

Keeping corrections of the order of $\delta u_{0\pm}/\gamma_0$ and $a_0^2/\gamma_0^2$, the Lorentz factor can be presented as
\begin{equation}
\label{eq:gammafast}
\gamma_\pm = \gamma_0 - \frac{u_0}{\gamma_0}\delta u_{0z\pm} + \frac{\eta_\pm^2\left|\bm{a}_0\right|^2}{4\gamma_0} + \frac{\eta_\pm^2\bm{a}_0^2}{8\gamma_0} {\rm e}^{-2{\rm i}\omega t} + {\rm c.c.}\;,
\end{equation}
where we neglected a term proportional to $a_{0z}$. Since the Coulomb gauge implies $a_{0z}/l_z\sim a_{0\perp}/l_\perp$, the longitudinal size of the wave packet should be shorter than its transverse size (i.e.,~$l_z\ll l_\perp$ in the wave frame, which is satisfied when the radio burst illuminates a substantial fraction of the solid angle).

The fast oscillations of the proper number density can be determined by substituting Eq.~\eqref{eq:gammafast} into Eq.~\eqref{eq:cont}. Neglecting the spatial derivatives, one finds
\begin{equation}
\label{eq:nfast}
n_\pm = n_0 + \delta n_{0\pm} - \frac{\eta_\pm^2\bm{a}_0^2n_0}{8\gamma_0^2} {\rm e}^{-2{\rm i}\omega t} + {\rm c.c.}\;,
\end{equation}
where $\delta n_{0\pm}$ describes the secular evolution of the proper density. We will keep corrections of the order of $\delta n_{0\pm}/n_0$.

\subsection{Secular evolution}
\label{sec:secular}

Now we derive the equations that govern the secular evolution of the physical quantities. Here retaining the spatial derivatives is crucial. For example, the gradient of the radiation intensity affects the dynamics of the plasma via the ponderomotive force \cite{Kruer2019}. We refer the reader who is not interested in the details of the calculation to our final results, which are Eqs.~\eqref{eq:final1}, \eqref{eq:final2}, \eqref{eq:final3}.

The secular evolution of the four-velocity can be determined substituting Eqs.~\eqref{eq:a}-\eqref{eq:gammafast} into Eq.~\eqref{eq:motion}. Averaging over the fast oscillations, one finds
\begin{align}
\nonumber
\frac{\pa}{\pa t} & \left(\delta\bm{u}_{0\pm} \pm \eta_\pm\bm{\psi}_0 \right) +\bm{e}_z\times\left[\nabla\times\left(\delta\bm{u}_{0\pm} \pm \eta_\pm\bm{\psi}_0 \right)\right] = \\
\label{eq:uslow}
& = \nabla\left(\delta u_{0z\pm} - \frac{\eta_\pm^2\left|\bm{a}_0\right|^2}{4\gamma_0} \mp \eta_\pm\phi_0\right) \;,
\end{align}
where $\phi_0$ describes the secular evolution of the scalar potential. In the derivation of Eq.~\eqref{eq:uslow}, we approximated $u_0/\gamma_0\simeq 1$, as appropriate because $\gamma_0 \gg 1$. It is convenient to define the variables $\gamma_0(1+\eta_+)\delta\bm{v}=\delta\bm{u}_{0+}+\eta_+\delta\bm{u}_{0-}$ and $\gamma_0(1+\eta_+)\delta\bm{q} =\delta\bm{u}_{0+}-\delta\bm{u}_{0-} + (1+\eta_+)\bm{\psi}_0$, which are governed by
\begin{align}
\label{eq:uslow1}
\frac{\pa\delta\bm{v}}{\pa t} - \frac{\pa\delta\bm{v}}{\pa z}  & = - \eta_+ \nabla\frac{\left|\bm{a}_0\right|^2}{4\gamma_0^2} \\
\label{eq:uslow2}
\frac{\pa\delta\bm{q}}{\pa t} - \frac{\pa\delta\bm{q}}{\pa z} & = - \nabla\left[ \frac{\psi_{0z}+\phi_0}{\gamma_0} - \frac{\left(1-\eta_+\right)\left|\bm{a}_0\right|^2}{4\gamma_0^2} \right] \;.
\end{align}
In the derivation of Eqs.~\eqref{eq:uslow1}-\eqref{eq:uslow2}, we used the identity $\bm{e}_z\times(\nabla\times\delta\bm{v})=\nabla_\perp \delta v_z-\pa\delta\bm{v}_\perp/\pa z$. The right hand side of Eq.~\eqref{eq:uslow1} is the ponderomotive force.

The secular evolution of the proper number density can be determined by substituting Eqs.~\eqref{eq:ufast}-\eqref{eq:nfast} into Eq.~\eqref{eq:cont}, and averaging over the fast oscillations. It is convenient to introduce a new variable, $(1+\eta_+)\delta \rho = \delta n_{0+}/n_0+\eta_+\delta n_{0-}/n_0-(1+\eta_+) \delta v_z$, which is governed by
\begin{equation}
\label{eq:nslow2}
\frac{\pa\delta \rho}{\pa t} +\eta_+\frac{\pa}{\pa t}\frac{\left|\bm{a}_0\right|^2}{4\gamma_0^2} - \frac{\pa\delta \rho}{\pa z} =  -\nabla_\perp\cdot\delta\bm{v}_\perp \;.
\end{equation}

Now we study the secular evolution the electromagnetic fields, and derive Eq.~\eqref{eq:gaussslow2}. Substituting Eqs.~\eqref{eq:a}, \eqref{eq:ufast}, \eqref{eq:nfast} into Eq.~\eqref{eq:maxwell1}, and averaging over the fast oscillations, one finds
\begin{align}
\nonumber
\frac{1+\eta_+}{\gamma_0\omega_{\rm P}^2} & \left(\frac{\pa^2\bm{\psi}_0}{\pa t^2}-\nabla^2\bm{\psi}_0 +\nabla\frac{\pa\phi_0}{\pa t} \right) +\\
\nonumber
&  -\frac{1}{2\gamma_0^2}\left(\frac{\delta n_{0+}}{n_0} - \frac{\delta n_{0-}}{n_0}  \right)\bm{e}_z= \\
\label{eq:ampereslow0}
& = \frac{\delta\bm{u}_{0+}}{\gamma_0} - \frac{\delta\bm{u}_{0-}}{\gamma_0} - \left(\frac{\delta n_{0+}}{n_0} - \frac{\delta n_{0-}}{n_0}\right)\bm{e}_z \;.
\end{align}
In the derivation of Eq.~\eqref{eq:ampereslow0}, we approximated $u_0/\gamma_0\simeq 1-1/2\gamma_0^2$. Here one should keep corrections of the order of $1/\gamma_0^2$ because the leading terms cancel, as we show below. Since the average physical quantities vary on a scale comparable to the length of the wave packet, which in the wave frame is much longer than $1/\omega_{\rm P}$, one has, for example, $\nabla^2\bm{\psi}_0/\omega_{\rm P}^2\ll\bm{\psi}_0$. Then, the terms on the right hand side of Eq.~\eqref{eq:ampereslow0} should nearly balance, as they are much larger than the terms on the left hand side. The transverse and longitudinal components of Eq.~\eqref{eq:ampereslow0} give respectively $\delta\bm{q}_\perp=\bm{\psi}_{0\perp}/\gamma_0$, and
\begin{equation}
\label{eq:nslow1}
\frac{\delta n_{0+}}{n_0} - \frac{\delta n_{0-}}{n_0} = \left(1+\eta_+\right)\left(\delta q_z-\frac{\psi_{0z}}{\gamma_0}\right) \;.
\end{equation}
Substituting Eqs.~\eqref{eq:gammafast}-\eqref{eq:nfast} into Eq.~\eqref{eq:maxwell0}, and averaging over the fast oscillations, one finds
\begin{align}
\nonumber
& \frac{1+\eta_+}{\gamma_0\omega_{\rm P}^2} \nabla^2\phi_0 + \frac{1}{2\gamma_0^2}\left(\frac{\delta u_{0z+}}{\gamma_0} - \frac{\delta u_{0z-}}{\gamma_0}  \right) + \\
\label{eq:gaussslow0}
& - \frac{\left(1-\eta_+^2\right)\left|\bm{a}_0\right|^2}{4\gamma_0^2} = \frac{\delta u_{0z+}}{\gamma_0} - \frac{\delta u_{0z-}}{\gamma_0} - \frac{\delta n_{0+}}{n_0} + \frac{\delta n_{0-}}{n_0}  \;.
\end{align}
One should combine the longitudinal component of Eq.~\eqref{eq:ampereslow0} and Eq.~\eqref{eq:gaussslow0} to eliminate the large terms on the right hand sides. Then, one can use the approximate expression given by Eq.~\eqref{eq:nslow1}. This procedure gives
\begin{align}
\nonumber
\frac{1}{\omega_{\rm P}^2} & \left(\frac{\pa^2}{\pa t^2}\frac{\psi_{0z}}{\gamma_0}-\nabla^2\frac{\psi_{0z}+\phi_0}{\gamma_0} + \frac{\pa^2}{\pa t\pa z} \frac{\phi_0}{\gamma_0}\right) +\\
\label{eq:gaussslow1}
& -\frac{1}{\gamma_0^2}\left(\delta q_z-\frac{\psi_{0z}}{\gamma_0}\right) + \frac{\left(1-\eta_+\right)\left|\bm{a}_0\right|^2}{4\gamma_0^2} = 0 \;.
\end{align}
In order to eliminate the spatial derivatives in the transverse direction, one should substitute $\delta\bm{q}_\perp=\bm{\psi}_{0\perp}/\gamma_0$ into the transverse component of Eq.~\eqref{eq:uslow2}, take the divergence of both sides, and use the gauge condition $\nabla_\perp\cdot\bm{\psi}_{0\perp}=-\pa\psi_{0z}/\pa z$. Then, isolate the term $\nabla^2_\perp (\psi_{0z}+\phi_0)/\gamma_0$, and substitute it into Eq.~\eqref{eq:gaussslow1}. This procedure gives
\begin{align}
\nonumber
\frac{1}{\omega_{\rm P}^2} & \left(\frac{\pa^2}{\pa t^2}\frac{\psi_{0z}}{\gamma_0}-\frac{\pa^2}{\pa z^2}\frac{\phi_0}{\gamma_0} + \frac{\pa^2}{\pa t\pa z} \frac{\phi_0-\psi_{0z}}{\gamma_0}\right) +\\
\label{eq:gaussslow2}
& -\frac{1}{\gamma_0^2}\left(\delta q_z-\frac{\psi_{0z}}{\gamma_0}\right) + \frac{\left(1-\eta_+\right)\left|\bm{a}_0\right|^2}{4\gamma_0^2} = 0 \;.
\end{align}

The equations that govern the secular evolution of the physical quantities (i.e.,~Eqs.~\ref{eq:uslow1}, \ref{eq:uslow2}, \ref{eq:nslow2}, \ref{eq:gaussslow2}) can be significantly simplified by neglecting the time derivative of $|\bm{a}_0|^2$. This approximation is justified because we work in the frame that moves with the group velocity of the wave, where the time evolution of the wave envelope is slow.\footnote{Our approximation is similar to the classical quasistatic approximation \cite{Sprangle+1990, Sprangle+1990b, Sprangle+1992}. However, we retain the time derivative of the fluid variables in Eqs.~\eqref{eq:final1}-\eqref{eq:final2}, and the time derivative of the electromagnetic fields in Eq.~\eqref{eq:final3}. As we show below, this is crucial to study the instabilities.} More formally, one can show that $\pa |\bm{a}_0|^2/\pa t\ll \nabla|\bm{a}_0|^2$ from Eqs.~\eqref{eq:final4ee}-\eqref{eq:final4ei} below. In this approximation, the longitudinal component of Eq.~\eqref{eq:uslow1} gives $\delta v_z=\eta_+|\bm{a}_0|^2/4\gamma_0^2$. The transverse component of Eq.~\eqref{eq:uslow1} and Eq.~\eqref{eq:nslow2} give respectively
\begin{align}
\label{eq:final1}
\frac{\pa\delta s}{\pa t} & - \frac{\pa\delta s}{\pa z} = - \eta_+ \nabla_\perp^2 \frac{\left|\bm{a}_0\right|^2}{4\gamma_0^2} \\
\label{eq:final2}
\frac{\pa\delta \rho}{\pa t} & - \frac{\pa\delta \rho}{\pa z} =  -\delta s \;,
\end{align}
where $\delta s=\nabla_\perp\cdot\delta\bm{v}_\perp$.

One cannot neglect the time derivatives in Eqs.~\eqref{eq:uslow2} and \eqref{eq:gaussslow2}. The reason becomes clear after a Fourier transform. Defining $\delta q_z = \int{\rm d}^3K{\rm d}\Omega\;\delta\tilde{q}_z \exp[{\rm i}(\bm{K}\cdot\bm{x}-\Omega t)]$, the longitudinal component of Eq.~\eqref{eq:uslow2} gives
\begin{equation}
\label{eq:qz}
\delta q_z= \frac{K_z}{K_z+\Omega} \left[\frac{\psi_{0z}+\phi_0}{\gamma_0} - \frac{\left(1-\eta_+\right) \left|\bm{a}_0\right|^2}{4\gamma_0^2} \right] \;,
\end{equation}
where we dropped the tilde. Since the time evolution is slow, one may neglect the second order time derivatives, and approximate $K_z/(K_z+\Omega)\simeq 1-\Omega/K_z$ in Eq.~\eqref{eq:qz}. Substituting the resulting expression into the Fourier transform of Eq.~\eqref{eq:gaussslow2}, and neglecting the second order time derivative of $\psi_{0z}$, one finds
\begin{align}
\nonumber
\left(\frac{1}{\gamma_0^2} \right. & - \left. \frac{K_z^2}{\omega_{\rm P}^2}\right) \frac{\phi_0}{\gamma_0} -\frac{\Omega K_z}{\omega_{\rm P}^2} \left(1+\frac{\omega_{\rm P}^2}{\gamma_0^2 K_z^2}\right)\frac{\phi_0}{\gamma_0} + \\
\label{eq:gaussslowFT}
& + \frac{\Omega K_z}{\omega_{\rm P}^2} \left(1-\frac{\omega_{\rm P}^2}{\gamma_0^2 K_z^2}\right)\frac{\psi_{0z}}{\gamma_0} = \frac{\left(1-\eta_+\right)\left|\bm{a}_0\right|^2}{4\gamma_0^2} \;.
\end{align}
In the derivation of Eq.~\eqref{eq:gaussslowFT}, we neglected terms of the order of $a_0^2/\gamma_0^4$, which are much smaller than the term on the right hand side of Eq.~\eqref{eq:gaussslowFT} because $\gamma_0\gg 1$.

Substituting $\Omega=0$ into Eq.~\eqref{eq:gaussslowFT} would be incorrect near the resonance $K_z^2=\omega_{\rm P}^2/\gamma_0^2$, where the first term on the left hand side vanishes. Instead, one should substitute $K_z^2=\omega_{\rm P}^2/\gamma_0^2$ into the second and third terms. Then, the inverse Fourier transform of Eq.~\eqref{eq:gaussslowFT} gives
\begin{equation}
\label{eq:final3}
\frac{2}{\omega_{\rm P}^2}\frac{\pa^2}{\pa t\pa z} \frac{\phi_0}{\gamma_0} = \frac{1}{\omega_{\rm P}^2} \frac{\pa^2}{\pa z^2}\frac{\phi_0}{\gamma_0} + \frac{1}{\gamma_0^2}\frac{\phi_0}{\gamma_0} - \frac{\left(1-\eta_+\right)\left|\bm{a}_0\right|^2}{4\gamma_0^2} \;.
\end{equation}
In pair plasmas ($\eta_+=1$) one has $\phi_0=0$, whereas in electron-proton plasmas ($\eta_+\ll 1$) a non-vanishing average electric field is generated. Since the time evolution is slow, the terms on the right hand side should nearly balance. Then, Eq.~\eqref{eq:final3} is analogous to a forced harmonic oscillator. The solutions of the homogeneous equation are plasma waves. In the lab frame, the phase velocity of the plasma waves is equal to the group velocity of the electromagnetic pulse \cite{TajimaDawson1979}.

When $\tau_0\gg 1/\omega_{\rm P}$, where $\tau_0=l_z/\gamma_0$ is the pulse duration in the lab frame, Eq.~\eqref{eq:final3} has the approximate solution $\phi_0/\gamma_0=|\bm{a}_0|^2/4$. There is no wakefield, as $\phi_0=0$ in the region behind the pulse where $\bm{a}_0=0$. In contrast, when $\tau_0\ll 1/\omega_{\rm P}$ the forcing term proportional to $|\bm{a}_0|^2$ is practically impulsive. The strength of the potential within the pulse is $\phi_0/\gamma_0 \sim a_0^2 \omega_{\rm P}^2\tau_0^2$, and reaches its maximum in the tail of the pulse. The strength of the potential in the wake is equal to the value of the derivative immediately behind the pulse, $(\pa/\pa z)(\phi_0/\gamma_0)\sim a_0^2 \omega_{\rm P}^2\tau_0/\gamma_0$, times the period of the plasma wave, $\gamma_0/\omega_{\rm P}$, which gives $\phi_0/\gamma_0\sim a_0^2\omega_{\rm P}\tau_0$. In short pulses where $\tau_0\ll 1/\omega_{\rm P}$, the wakefield potential is much larger (by a factor of $1/\omega_{\rm P}\tau_0$) than the maximal potential within the pulse.

\subsection{When are nonlinear effects weak?}

Our results are valid when the particle Lorentz factor is nearly constant in the wave frame, so that nonlinear effects are weak. Eq.~\eqref{eq:gammafast} shows that the condition $\gamma_\pm\simeq\gamma_0$ requires $a_0\ll\gamma_0$ and $\delta u_{0z\pm}\ll\gamma_0$. Since $\delta u_{0z\pm}= \pm\eta_\pm\phi_0+\eta_\pm^2|\bm{a}_0|^2/4\gamma_0$, one needs $a_0\ll\gamma_0$ and $\phi_0\ll\gamma_0$. In pair plasmas ($\eta_+=1$), one has $\phi_0=0$. Then, nonlinear effects are weak for\footnote{Kennel and Pellat studied the propagation of plane waves in pair plasmas \cite{KennelPellat1976}. They found that the plane wave is a sine function for $a_0\ll\gamma_0$, whereas it has a sawtooth profile for $a_0\gg\gamma_0$. Plane wave solutions in the strongly nonlinear regime $a_0\gg\gamma_0$ may have limited practical importance, as particles would be trapped within a pulse of finite size.} $a_0\ll\gamma_0=\omega_0/\omega_{\rm P}$. When $a_0\gtrsim\gamma_0$, particles are trapped within the pulse, as their longitudinal velocity in the lab frame becomes comparable with the group velocity of the wave \cite{Schroeder+2006}.

In electron-proton plasmas ($\eta_+\ll 1$), nonlinear effects are weak in the nonrelativistic limit $a_0\ll 1$. The duration of the pulse is crucial when $a_0\gg 1$. For short pulses with $\tau_0\ll 1/\omega_{\rm P}$, the strength of the potential is $\phi_0/\gamma_0 \sim a_0^2\omega_{\rm P}^2\tau_0^2$ within the pulse, and $\phi_0/\gamma_0 \sim a_0^2\omega_{\rm P}\tau_0$ in the wake (these estimates are valid for $\phi_0\ll\gamma_0$). When $\tau_0\ll 1/a_0^2\omega_{\rm P}$, our model describes both the pulse and the wakefield. When $1/a_0^2\omega_{\rm P}\ll \tau_0\ll 1/a_0\omega_{\rm P}$, the condition $\phi_0\ll\gamma_0$ is satisfied within the pulse, whereas the wakefield is strongly nonlinear.\footnote{Sprangle, Esarey, and collaborators studied the pulse propagation in the limit $\eta_+=0$ \cite{Sprangle+1990, Sprangle+1990b, Sprangle+1992}. Relativistic optical guiding of short pulses was found to be suppressed when $\tau_0\ll 1/a_0\omega_{\rm P}$, in agreement with our results.} Electrons trapped in the nonlinear plasma wave can be accelerated to Lorentz factors $\gtrsim\gamma_0^2$ in the lab frame \cite{TajimaDawson1979, EsareyPilloff1995}. Our model describes the evolution of the pulse, but not the wakefield acceleration. When $\tau_0\gg 1/a_0\omega_{\rm P}$, our model describes only the leading portion of the pulse, of duration $\sim 1/a_0\omega_{\rm P}$, as the condition $\phi_0\ll\gamma_0$ is violated in the tail of the pulse.

\section{Evolution of the wave envelope}
\label{sec:envelope}

Eqs.~\eqref{eq:final1}, \eqref{eq:final2}, \eqref{eq:final3} should be complemented with the equation that governs the evolution of the wave envelope. Substituting Eqs.~\eqref{eq:a}, \eqref{eq:ufast}, \eqref{eq:nfast} into Eq.~\eqref{eq:maxwell1}, and considering the resonant terms (i.e., the terms proportional to ${\rm e}^{-{\rm i}\omega t}$), one finds
\begin{align}
\nonumber
2 {\rm i}\omega & \frac{\pa\bm{a}_0}{\pa t} = \left(\omega_{\rm P}^2-\omega^2\right)\bm{a}_0 - \nabla^2\bm{a}_0 + \\
\label{eq:env}
& + \omega_{\rm P}^2 \left[\delta \rho -\left(1-\eta_+\right)\frac{\phi_0}{\gamma_0} + \frac{1+\eta_+^3}{1+\eta_+} \frac{\left|\bm{a}_0\right|^2}{8\gamma_0^2} \right] \bm{a}_0 \;.
\end{align}
In the derivation of Eq.~\eqref{eq:env}, we used the relations $\delta n_{0+}/n_0+\eta_+\delta n_{0-}/n_0= (1+\eta_+)[\delta \rho+\eta_+|\bm{a}_0|^2/4\gamma_0^2]$ and $\delta n_{0+}/n_0-\delta n_{0-}/n_0=(1+\eta_+)[\phi_0/\gamma_0-(1-\eta_+)|\bm{a}_0|^2/4\gamma_0^2]$, where the latter relation follows from Eqs.~\eqref{eq:nslow1} and \eqref{eq:qz}. We also neglected the second order time derivative of $\bm{a}_0$, as appropriate because the time evolution of the wave envelope is slow.

The first term on the right-hand side of Eq.~\eqref{eq:env} is much larger than the other terms, and therefore should vanish. This condition gives the dispersion relation of the wave, which is $\omega^2=\omega_{\rm P}^2$, or equivalently $\omega_0^2=k_0^2+\omega_{\rm P}^2$ in the lab frame. This implies $\gamma_0=\omega_0/\omega_{\rm P}$.

Eq.~\eqref{eq:env} can be simplified considering specific cases. In pair plasmas ($\eta_+=1$), one has
\begin{equation}
\label{eq:final4ee}
\frac{\rm i}{\omega_{\rm P}}\frac{\pa\bm{a}_0}{\pa t} = -\frac{1}{2\omega_{\rm P}^2}\nabla^2\bm{a}_0+\frac{1}{2}\delta \rho \bm{a}_0 + \frac{\left|\bm{a}_0\right|^2}{16\gamma_0^2} \bm{a}_0 \;.
\end{equation}
In electron-proton plasmas ($\eta_+\ll 1$), one has
\begin{equation}
\label{eq:final4ei}
\frac{\rm i}{\omega_{\rm P}}\frac{\pa\bm{a}_0}{\pa t} = -\frac{1}{2\omega_{\rm P}^2}\nabla^2\bm{a}_0+\frac{1}{2}\delta \rho \bm{a}_0 -\frac{1}{2}\frac{\phi_0}{\gamma_0}\bm{a}_0 \;.
\end{equation}
In the derivation of Eq.~\eqref{eq:final4ei}, we took into account that $\phi_0/\gamma_0\gg a_0^2/\gamma_0^2$. This condition is satisfied because the strength of the scalar potential within the pulse is $\phi_0/\gamma_0 \sim \min[a_0^2, \omega_{\rm P}^2a_0^2\tau_0^2]$. Since in the lab frame the pulse is longer than a few wavelengths, one has $\gamma_0\omega_{\rm P}\tau_0=\omega_0\tau_0\gg 1$, which implies $\omega_{\rm P}^2a_0^2\tau_0^2\gg a_0^2/\gamma_0^2$.

\subsection{Instabilities}

The effect of modulation/filamentation instabilities and stimulated scattering processes has been originally considered for laser plasma interaction \cite{Drake+1974, Max+1974, Forslund+1975, MckinstrieBingham1992, AntonsenMora1993}, and subsequently for pulsar radio emission \cite{BlandfordScharlemann1976, WilsonRees1978, GedalinEichler1993, Thompson+1994} and fast radio bursts \cite{Lyubarsky2008, Lyubarsky2019, Sobacchi+2021, Ghosh+2022, Sobacchi+2022, Sobacchi+2023}. These instabilities can be studied within our framework. One should find a plane wave solution of Eqs.~\eqref{eq:final1}, \eqref{eq:final2}, \eqref{eq:final3}, \eqref{eq:final4ee}-\eqref{eq:final4ei}, and then study the evolution of a small perturbation of the wave intensity. We discuss the limiting cases $\eta_+=1$ and $\eta_+=0$. Previous studies focusing on the nonrelativistic limit $a_0\ll 1$ suggest that a kinetic model would give the same growth rate of the instabilities as our fluid model \cite{Drake+1974, Ghosh+2022}.

In pair plasmas ($\eta_+=1$), one has $\phi_0=0$. Then, the evolution of the wave envelope is governed by Eqs.~\eqref{eq:final1}, \eqref{eq:final2}, \eqref{eq:final4ee}. We study the evolution of a small perturbation of the wave intensity by defining
\begin{equation}
\label{eq:apertee}
\bm{a}_0= \hat{\bm{n}}\left(1+\delta a\right) a_0 \exp\left[-{\rm i}\frac{a_0^2}{16\gamma_0^2}\omega_{\rm P}t\right] \;,
\end{equation}
where $\delta a\ll 1$, and $\hat{\bm{n}}$ is a unit vector (i.e.,~$|\hat{\bm{n}}|=1$). The dispersion relation can be determined by substituting Eq.~\eqref{eq:apertee} into Eqs.~\eqref{eq:final1}, \eqref{eq:final2}, \eqref{eq:final4ee}, and neglecting terms of the order of $(\delta a)^2$ (note that $\delta a=\delta s=\delta \rho=0$ is an exact solution of these equations). It is convenient to use the variables $\delta a+\delta a^*$ and $\delta a-\delta a^*$, where $\delta a^*$ is the complex conjugate of $\delta a$. Assuming that these variables are proportional to $\exp[{\rm i}(\bm{ K}\cdot\bm{ x}-\Omega t)]$, one finds
\begin{equation}
\label{eq:DRfinalee}
\left(\Omega+K_z\right)^2 \left( \frac{4\Omega^2}{K^2}-\frac{K^2}{\omega_{\rm P}^2}-\frac{a_0^2}{4\gamma_0^2}\right) = \frac{a_0^2K_\perp^2}{2\gamma_0^2}\;.
\end{equation}
Substituting $\Omega=-K_z+\Delta\Omega$ into Eq.~\eqref{eq:DRfinalee}, and considering long wavelengths in the longitudinal direction (i.e., $K_z\ll K_\perp$ and $K_z\ll |\Delta\Omega|$), one finds the growth rate of the filamentation instability. The maximal rate is given by $(\Delta\Omega)^2 = -(a_0^2/2\gamma_0^2)\omega_{\rm P}^2$, and it is achieved for wave numbers $K_\perp^2\gg (a_0/\gamma_0)\omega_{\rm P}^2$. The instability develops if the duration of the radiation pulse is longer than the inverse of the growth rate. When $|\Delta\Omega| l_z\gg 1$, or equivalently $a_0\omega_{\rm P}\tau_0\gg 1$, the pulse is broken into filaments parallel to the direction of propagation. Our framework cannot be used to study the saturation of the instability, as the density fluctuations become large \citep{Kaw+1973, Sun+1987, Iwamoto+2023}.

In the limit of infinitely massive ions ($\eta_+=0$), one has $\delta s=\delta \rho=0$. Then, the evolution of the wave envelope is governed by Eqs.~\eqref{eq:final3} and \eqref{eq:final4ei}. We study the evolution of a small perturbation of the wave intensity by defining
\begin{equation}
\label{eq:apertei}
\bm{a}_0= \hat{\bm{n}}\left(1+\delta a\right) a_0 \exp\left[{\rm i}\frac{a_0^2}{8}\omega_{\rm P}t\right] \;,
\end{equation}
and $\phi_0/\gamma_0=(a_0^2/4)(1+\delta\phi)$, where $\delta a\ll 1$ and $\delta\phi\ll 1$ (this solution is valid for $a_0\ll 1$). The derivation of the dispersion relation is analogous to the case of a pair plasma. One finds
\begin{equation}
\label{eq:DRfinalei}
\left(1-\frac{\gamma_0^2K_z^2}{\omega_{\rm P}^2} - \frac{2\gamma_0^2K_z\Omega}{\omega_{\rm P}^2} \right) \left(\frac{4\Omega^2}{K^2}-\frac{K^2}{\omega_{\rm P}^2} \right) = - \frac{a_0^2}{2} \;.
\end{equation}
The filamentation instability ($K_z=0$) develops for $K_\perp^2<a_0^2\omega_{\rm P}^2/2$. The maximal growth rate is given by $\Omega^2 = -a_0^4\omega_{\rm P}^2/64$, and it is achieved for $K_\perp^2=a_0^2\omega_{\rm P}^2/4$. Consider a plasma shell of width $\Delta R$ in the lab frame. The instability develops if the shell crossing time in the wave frame, $\Delta R/\gamma_0$, is longer than the inverse of the growth rate. When $|\Omega|\Delta R/\gamma_0 \gg 1$, or equivalently $a_0^2\omega_{\rm P}^2\Delta R/\omega_0\gg 1$, the radiation pulse is broken into filaments of transverse size $\sim 1/a_0\omega_{\rm P}$.

The modulational instability ($K_\perp=0$) can develop in electron-proton plasmas. When $a_0\omega_0\ll \omega_{\rm P}$, wave numbers $K_z^2<a_0^2\omega_{\rm P}^2/2$ are unstable, and the maximal growth rate is the same of the filamentation instability. In the lab frame, the longitudinal size of the radiation intensity modulations is $\sim 1/a_0\omega_0$. When $a_0\omega_0\gg \omega_{\rm P}$, the growth rate is given by $\Omega^2=-a_0^2K_z^2/8$, and large wave numbers ($K_z^2 > \omega_{\rm P}^2/\gamma_0^2$) are stabilized. The modulational instability merges with stimulated Raman scattering, which develops for $K_z^2 = \omega_{\rm P}^2/\gamma_0^2$.

\section{Discussion}
\label{sec:disc}

We introduced a framework to study the propagation of strong electromagnetic waves in tenuous plasmas where $\omega_{\rm P}\ll\omega_0$. We showed that in pair plasmas nonlinear effects are weak for $a_0\ll \omega_0/\omega_{\rm P}$. Instead, in electron-proton plasmas one needs either $a_0\ll 1/\omega_{\rm P}\tau_0$, where $\tau_0$ is the duration of the radiation pulse in the lab frame, or $a_0\ll 1$. In the weakly nonlinear regime, the evolution of the wave envelope is governed by Eqs.~\eqref{eq:final1}, \eqref{eq:final2}, \eqref{eq:final3}, and \eqref{eq:final4ee}-\eqref{eq:final4ei}.

Our results have important implications for the modeling of fast radio bursts (FRBs). For a typical luminosity $L\sim 10^{42}{\rm\; erg\; s^{-1}}$, and frequency $\nu_0=\omega_0/2\pi\sim 1{\rm\; GHz}$, the FRB strength parameter is $a_0\sim 200\; R_{11}^{-1}$, where $R=10^{11}R_{11}{\rm\; cm}$ is the distance from the source (most likely a magnetar) \cite{Lyubarsky2021, Zhang2023}. The composition of the plasma surrounding the magnetar is uncertain. A relativistic wind is formed outside the light cylinder, which is located at the distance $R\sim 10^{10}{\rm\; cm}$ assuming a rotation period of a few seconds. Magnetar winds are magnetized, and likely composed of electron-positron pairs \cite{KaspiBeloborodov2017}. On the other hand, magnetar flares can eject a significant amount of protons with mildly relativistic speeds \cite{Gaensler+2005, Gelfand+2005, Granot+2006}. The ejected electron-proton shell becomes weakly magnetized as it expands \cite{Beloborodov2020}. Below we argue that the composition of the plasma surrounding the source affects the observed properties of FRBs.

FRBs may be unable to propagate on astrophysically relevant scales when nonlinear effects are strong, as particles can be trapped within the pulse. Then, the kinetic energy of the trapped particles would eventually exceed the pulse electromagnetic energy. A smooth pulse of duration $\tau_0\sim 1{\rm\; ms}$ should be produced in a proton-free environment, as the condition $a_0\lesssim 1/\omega_{\rm P}\tau_0$ would require an unrealistically small density, $n_0\lesssim 10^{-8} R_{11}^2{\rm\; cm^{-3}}$.

An electron-proton shell near the source may affect the FRB time structure. Nonlinear effects are weak in the leading portion of the pulse, of typical duration $1/a_0\omega_{\rm P}\sim 90\; n_0^{-1/2}R_{11}{\rm\; ns}$. We envision two alternative scenarios. (i) The leading portion detaches from the rest of the pulse, which is therefore continuously eroded. This would produce a sequence of short pulses of duration $\sim 1/a_0\omega_{\rm P}$, consistent with the ultra-fast variability of some radio bursts \cite{Majid+2021, Nimmo+2021, Nimmo+2022}. (ii) The leading portion does not detach. The rest of the pulse may be unable to propagate on astrophysically relevant scales. This would produce a single short pulse, which could explain the recently discovered ultra-fast radio bursts \cite{Snelders+2023}. Fully-kinetic simulations can be used to study the propagation of FRBs in the strongly nonlinear regime.

Millisecond duration bursts with a smooth profile must be produced in a proton-free environment, where nonlinear effects are weaker. In unmagnetized pair plasmas, nonlinear effects are weak for $a_0\lesssim \omega_0/\omega_{\rm P}$, which requires $n_0\lesssim 10^5 R_{11}^2{\rm\; cm^{-3}}$. We will show elsewhere that when the magnetization is large (i.e.,~when $\omega_{\rm L}>\omega_{\rm P}$, where $\omega_{\rm L}$ is the Larmor frequency in the background magnetic field), nonlinear  effects are weak for $a_0\lesssim \omega_0/\omega_{\rm L}$. The latter condition implies that the magnetic field strength should be $B\lesssim 1\;R_{11}{\rm\; G}$. We will discuss elsewhere the implications of this constraint for FRB models.

\begin{acknowledgements}
This work was supported by the Marie Sk{\l}odowska-Curie Grant 101061217 [E.S.], by the JSPS KAKENHI Grants 20J00280, 20KK0064, and 22H00130 [M.I.], by the Simons Foundation Grant 00001470 to the Simons Collaboration on Extreme Electrodynamics of Compact  Sources (SCEECS) [L.S., T.P.], by the DoE Early Career Award DE-SC0023015 [L.S.], by the Multimessenger Plasma Physics Center (MPPC) NSF Grant PHY-2206609 [L.S.], by the ISF Grant 2126/22 [T.P.], and by the ERC Advanced Grant Multijets [T.P.]. We acknowledge insightful discussions with Andrei Beloborodov, Luca Comisso, Yuri Lyubarsky, and Chris Thompson.
\end{acknowledgements}

\bibliography{2d}

\end{document}